# A spin metal-oxide-semiconductor field-effect transistor using half-metallic-ferromagnet contacts for the source and drain

S. Sugahara[1,2] and M. Tanaka[1,2]
[1]*Dept. of Electronic Engineering, The University of Tokyo, 7-3-1 Hongo, Bunkyo-ku, Tokyo 113-8656, Japan*
[2]*PRESTO, Japan Science and Technology Agency, 4-1-8 Honcho, Kawaguchi, Saitama 332-0012, Japan*

We propose and theoretically analyze a novel metal-oxide-semiconductor field-effect-transistor (MOSFET) type of spin transistor (hereafter referred to as a spin MOSFET) consisting of a MOS gate structure and half-metallic-ferromagnet (HMF) contacts for the source and drain. When the magnetization configuration between the HMF source and drain is parallel (antiparallel), highly spin-polarized carriers injected from the HMF source to the channel are transported into (blocked by) the HMF drain, resulting in the magnetization-configuration-dependent output characteristics. Our two-dimensional numerical analysis indicates that the spin MOSFET exhibits high (low) current drive capability in the parallel (antiparallel) magnetization, and that extremely large magnetocurrent ratios can be obtained. Furthermore, the spin MOSFET satisfies other important requirements for "spintronic" integrated circuits, such as high amplification capability, low power-delay product, and low off-current.

Spin transistors, which utilize two ferromagnetic layers as a spin injector and a spin analyzer, possess unique output characteristics that are controlled by the relative magnetization configuration of the ferromagnets as well as the bias conditions[1-5]. Also, the magnetization configuration in spin transistors can be used as nonvolatile binary data. Owing to these useful features, spin transistors are potentially applicable to integrated circuits for ultrahigh-density nonvolatile memory whose memory cell is made of a single spin transistor[6] and for nonvolatile reconfigurable logic based on functional spin transistor gates[7]. In order to realize such "spintronic" integrated circuits with high performance, the following requirements must be satisfied for spin transistors; (i) large magnetocurrent ratio for nonvolatile memory and logic functions, (ii) high transconductance for high speed operation, (iii) high amplification capability (voltage, current and/or power gains) to restore propagating signals between transistors, (iv) small power-delay product and small off-current for low power dissipation, and (v) simple device structure for high degree of integration and high process yield.

Although various spin transistors have been proposed so far[1-5], none of them can satisfy all these requirements. Especially, the high transconductance and amplification capability cannot be realized simultaneously with the large magnetocurrent ratio. For example, large magnetocurrent ratios were reported in the spin-valve transistor proposed by Monsma et al[2] and spin transistors based on the similar operating principle[3,4], however it is difficult for these devices to achieve high transconductance and high power gain due to the existing tradeoff between their transfer ratio and magnetocurrent ratio. On the other hand, the spin field-effect-transistor (FET) proposed by Datta and Dass[5] can be expected to have high transconductance and high voltage gain, but its magnetocurrent ratio is limited to a very small value[8].

In this paper, we propose and theoretically analyze a novel metal-oxide-semiconductor FET (MOSFET) type of spin transistor, hereafter referred to as a spin MOSFET, consisting of a MOS gate structure and half-metallic-ferromagnet (HMF) contacts for the source and drain. The proposed spin MOSFET can simultaneously satisfy all the above-described requirements (i)-(v) for spintronic integrated circuits.

Figure 1(a) schematically shows the device structure of the proposed spin MOSFET that can be applied to not only $n$-channel and $p$-channel devices but also accumulation- and inversion-type channel devices. The structure of the spin MOSFET is similar to that of Schottky source/drain MOSFETs[9,10] except the HMF source/drain contacts that are the HMF/Si junctions without a $pn$ junction. Possible candidates for the HMF materials are Heusler alloys, $CrO_2$, $Fe_2O_3$ and ferromagnetic semiconductors[11-14]. Nonmagnetic (NM) contacts are also formed on the HMF source/drain (not shown in Fig. 1(a)). In the following, the $n$-channel accumulation-type spin MOSFET with an intrinsic Si channel layer is used to explain the operating principle of the spin MOSFET.

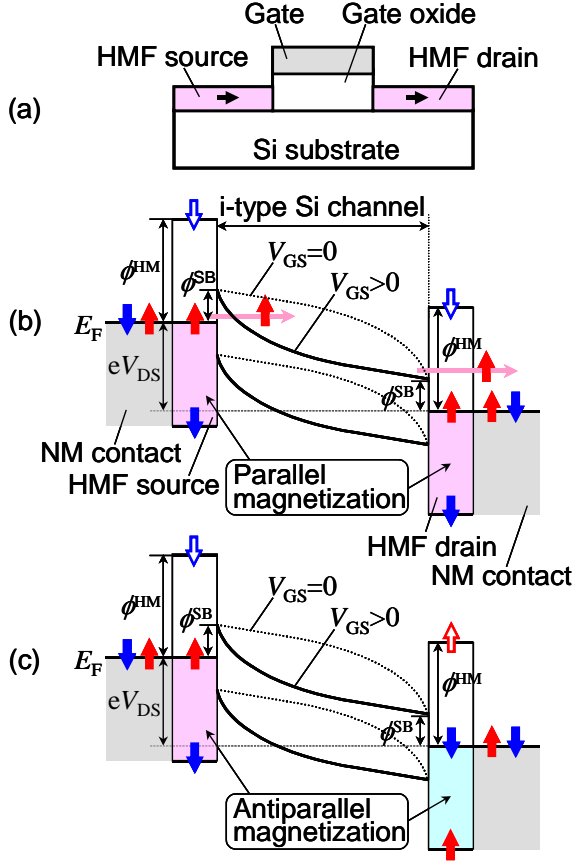

Fig. 1 Schematic (a) device structure and band diagrams of the spin MOSFET in (b) parallel and (c) antiparallel magnetization configurations.

Figure 1(b) schematically shows the band diagram of the spin MOSFET under a common source bias condition with and without a gate-source bias $V_{GS}$, where the relative magnetization configuration of the HMF source/drain is parallel. Owing to the metallic and insulating spin bands of the HMF source/drain material, spin-dependent barrier structures appear as shown in the figure, i.e., a Schottky barrier (SB) with a lower barrier height $\phi^{SB}$ for up-spin electrons and a rectangular energy barrier with a higher barrier height $\phi^{HM}$ for down-spin electrons (hereafter, this spin configuration at the HMF source is used throughout this paper). When a drain-source bias $V_{DS}$ (> 0) is applied with $V_{GS} = 0$, neither up-spin nor down-spin electrons are injected from the source to the channel due to the reverse-biased source-SB for up-spin electrons (as shown by the upper dotted curve in Fig. 1(b)) and the high rectangular barrier for down-spin electrons. By applying $V_{GS}$ (> 0), the width of the source-SB is reduced (as shown by the upper solid curve in the Fig. 1(b)) and thus up-spin electrons in the metallic spin band of the HMF source can tunnel through the thinned source-SB into the channel. On the other hand, the injection of down-spin electrons is blocked even under the application of $V_{DS}$ and $V_{GS}$ owing to the high rectangular barrier at the source. Thus, the HMF source acts not only as a contact for blocking an off-current but also as a spin-injector of up-spin electrons from the HMF source to the channel. In the parallel magnetization configuration, the up-spin electrons injected in the channel can be transported to the nonmagnetic drain contact through the metallic up-spin band of the HMF drain, resulting in a drain current. By flipping the magnetization of the HMF drain, the antiparallel spin configuration is established and the HMF barrier height for up-spin electrons becomes larger at the drain, as shown in Fig. 1(c). Thus, the up-spin electrons hardly pass through the HMF drain to the nonmagnetic drain contact. Namely, the HMF drain has the function of a spin-analyzer, i.e., the HMF drain selectively extracts the up-spin electrons from the channel when the magnetic configuration between the HMF source and drain is parallel. By combing these spin-filter effects of the HMF source/drain, an extremely large magnetocurrent ratio can be expected due to the high spin-selectivity of the HMF source/drain.

A model device used in our analysis is shown in Fig. 2(a), where a thin-film-transistor structure was used for the simplicity of calculation. The size of this model device is as follows; the gate oxide (SiO$_2$) thickness $t_{OX}$ is 2.0-3.0 nm, the Si layer thickness $t_{Si}$ is 10 nm and the channel length $L_{CH}$ is 30 nm. A device parameter $L_S$ ($L_D$) shown in the figure is the distance from the source (drain) junction to the nonmagnetic contact, which qualitatively represents the thickness of the HMF source/drain shown in Fig. 1(a). An intrinsic Si layer was used for the channel and ballistic transport was assumed for the spin-polarized electrons injected in the channel. A relatively small SB height of $\phi^{SB} = 0.2$ eV for the metallic spin band of the HMF source/drain was taken in order to achieve a large drain current. A barrier height of $\phi^{HM} = 1.0$ eV for the rectangular barrier of the HMF source/drain and a distance of $L_S = 5$ nm (= $L_D$) were selected in order to obtain the fully spin-polarized electron injection from the HMF source[15]. The effective mass $m_{Si}^*$ of the Si layer used in the calculation was 0.19 $m_0$, where $m_0$ is the free electron mass, and effective masses $m_M^* = m_0$ and $m_I^* = m_{Si}^*$ were assumed for the metallic and insulating spin bands of the HMF source/drain, respectively. The operating temperature was set at 300 K in all the calculations. Output

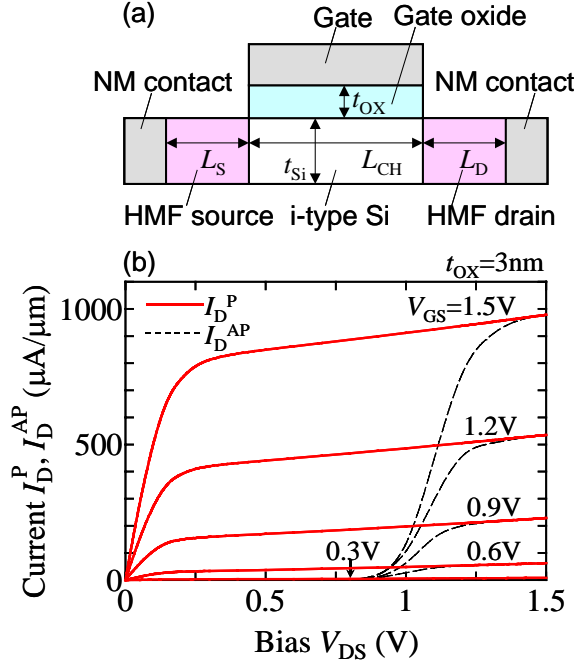

Fig. 2 (a) Device structure used for the analysis. The size of the device in our calculation is as follows; $t_{OX}$ = 2.0-3.0 nm, $t_{Si}$ = 10 nm, $L_{CH}$ = 30 nm, and $L_S$ = $L_D$ = 5 nm. (b) Output characteristics of the spin MOSFET. The drain currents $I_D^P$ (solid curves) and $I_D^{AP}$ (dashed curves) in the parallel and antiparallel magnetic configurations, respectively, are plotted as a function of $V_{DS}$, where $V_{GS}$ is varied from 0.3 to 1.5 V and $t_{OX}$ is 3.0 nm.

characteristics were calculated by using the Tsu-Esaki formula[16] with a two-dimensional transmission probability calculation. The detailed calculation procedure will be described elsewhere[15].

Solid and dashed curves in Fig. 2(b) show the calculated output characteristics of the spin MOSFET for the parallel and antiparallel magnetization configurations, respectively, where $t_{OX}$ is 3 nm. In the parallel magnetization, the drain current $I_D^P$ starts to increase at $V_{GS}$ more than 0.3 V (indicating the threshold voltage of 0.3 V) and increases nonlinearly with increasing $V_{GS}$, while $I_D^P$ shows saturation behavior for $V_{DS}$. This output characteristics can be explained by the bias-induced potential profile of the channel region as follows: When $V_{GS}$ and $V_{DS}$ are applied simultaneously, $V_{GS}$ induces a much stronger electric field from the gate electrode to the HMF source through the source-SB than that $V_{DS}$ does from the drain to the source. Thus, the width of the source-SB for tunneling emission is thinned by $V_{GS}$ as shown in Fig. 1(b) and it is insensitive to $V_{DS}$, resulting in the above described output characteristics. The value of $I_D^P$ is comparable to that of sub-100 nm scale MOSFETs[17], and $I_D^P$ increases with decreasing the SB height ($\phi^{SB}$) and gate oxide thickness ($t_{OX}$), like the conventional Schottky source/drain MOSFETs[10,18]. A large $I_D^P$ more than 1500 μA/μm can be obtained for $\phi^{SB}$ = 0.2 eV and $t_{OX}$ = 2.0 nm with a gate bias condition of $V_{GS}$ = 1.5 V. Note that reduction of $t_{OX}$ is also important to obtain the saturation behavior of $I_D^P$, and the channel conductance of the spin MOSFET (discussed later) is improved by the reduction of $t_{OX}$.

In the antiparallel magnetization configuration, the drain current $I_D^{AP}$ is negligibly small for $V_{DS}$ less than ($\phi^{HM}-\phi^{SB}$)/$e$ (= 0.8 V), but $I_D^{AP}$ increases exponentially with increasing $V_{DS}$ and reaches the same current value as $I_D^P$ when $V_{DS}$ is more than $\phi^{HM}/e$ (= 1.0 V), as shown in Fig. 2(b). Thus, magnetization-configuration-dependent output characteristics are realized when $V_{DS} < (\phi^{HM}-\phi^{SB})/e$. The exponential increase of $I_D^{AP}$ can be attributed to the ballistic transport in the channel region, i.e., up-spin electrons injected from the HMF source can pass over the large rectangular barrier of the HMF drain when $V_{DS}$ increases to more than ($\phi^{HM}-\phi^{SB}$)/$e$ = 0.8 V. Note that when the drain current is governed by drift-diffusion kinetics rather than ballistic transport, the exponential increase of $I_D^{AP}$ is significantly suppressed. In this case, however, dynamically accumulated spin-polarized electrons in the channel would affect $I_D^{AP}$ due to their finite spin lifetime for spin flipping. This effect can be treated by a self-consistent calculation based on a drift-diffusion model[18] including the spin lifetime, which will be reported elsewhere.

Figure 3(a) shows the magnetocurrent ratio $\gamma_{MC}$ of the spin MOSFET as a function of $V_{DS}$ at $V_{GS}$ = 1.5 V, where $\gamma_{MC}$ is defined by $(I_D^P-I_D^{AP})/I_D^{AP}$. $\gamma_{MC}$ exponentially falls with increasing $V_{DS}$, since $I_D^{AP}$ increases exponentially with increasing $V_{DS}$ when $V_{DS}$ is less than $\phi^{HM}/e$ as described above. In spite of this bias-dependence, extremely large $\gamma_{MC}$ more than 1000 % can be obtained for $V_{DS}$ less than 1.0 V. To obtain even larger $\gamma_{MC}$, higher $\phi^{HM}$ values are required. It should be noted that when the drift-diffusion transport is dominant, this strong bias-dependence of $\gamma_{MC}$ is significantly suppressed and extremely large $\gamma_{MC}$ will be obtained even at $V_{DS}$ = 1.5 V.

The results shown in Figs 2(b) and 3(a) indicate that the spin MOSFET possesses the magnetization-configuration-dependent output characteristics with large $\gamma_{MC}$. Thus, the spin MOSFET satisfies the above-mentioned requirement

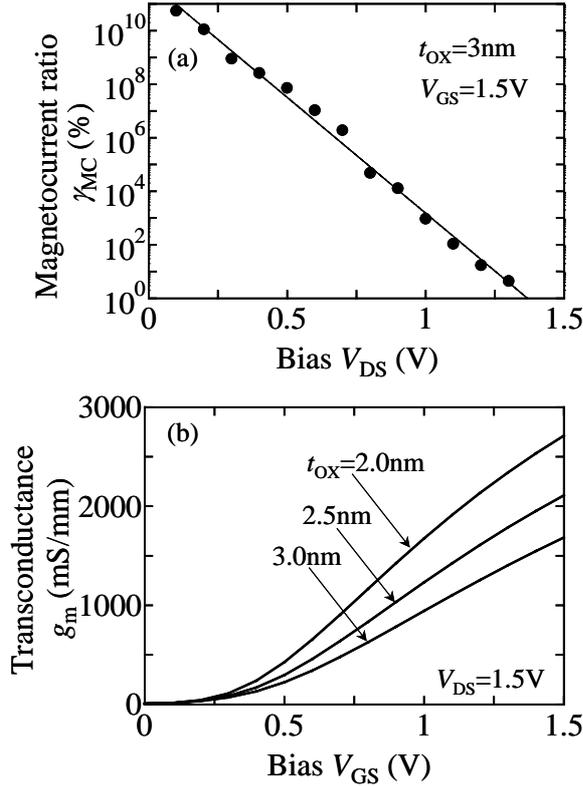

Fig. 3 (a) Magnetocurrent ratio $\gamma_{MC}$ [= $(I_D^P - I_D^{AP})/I_D^{AP}$] as a function of $V_{DS}$ at $V_{GS}$ = 1.5 V. (b) Transconductance $g_m$ (= $\partial I_D^P/\partial V_{GS}$) as a function of $V_{GS}$ at $V_{DS}$ = 1.5 V for $t_{OX}$ = 2.0, 2.5 and 3.0 nm.

(i) for spintronic integrated circuit applications. The other requirements were examined by the on- and off-current characteristics of the spin MOSFET as follows: Figure 3(b) shows the transconductance $g_m$ of the spin MOSFET in the parallel magnetization configuration as a function of $V_{GS}$ at $V_{DS}$ = 1.5 V, where $t_{OX}$ is varied from 2.0 to 3.0 nm. Here, $g_m$ is defined by a derivative $\partial I_D^P/\partial V_{GS}$ under a fixed $V_{DS}$ condition, which is a measure of the output-current ($I_D^P$) drive capability for the input voltage ($V_{GS}$). Since $I_D^P$ increases nonlinearly ($I_D^P \propto F^2\exp(-1/F)$) with increasing the strength $F$ of the electric field through the source-SB[10], $g_m$ increases with increasing $V_{GS}$ and with decreasing $t_{OX}$ as shown in Fig. 3(b). A large $g_m$ of 1000 mS/mm, which is comparable to (or larger than) that of sub-100 nm scale MOSFETs[17], can be obtained at $V_{GS}$ = 1.0 V for $t_{OX}$ = 3 nm, and $g_m$ is further enhanced to more than 1500 mS/mm by reducing $t_{OX}$ to 2 nm, as shown in the figure. These large $g_m$ values of the spin MOSFET lead to a small propagation delay $t_{pd}$ and a large voltage gain $G_V$, since $t_{pd}$ and $G_V$ can be estimated by $C_L/g_m$ and $g_m/g_D$, respectively, where $C_L$ is a load capacitance including a parasitic capacitance and $g_D$ is a channel conductance given by a derivative $\partial I_D^P/\partial V_{DS}$.

Furthermore, the large $g_m$ of the spin MOSFET enables low-voltage operation, and the voltage swing can be less than 1.0 V. This results in a small power-delay product $P \cdot t_{pd}$ (that corresponds to the energy per switching), since this energy is proportional to the square of the voltage swing[19]. The power dissipation is also caused by the off-current in the stand-by condition of the spin MOSFET, which is characterized by the subthreshold swing $S$ calculated from $\log I_D^P$-$V_{GS}$ characteristics. Although $S$ depends on $t_{OX}$ and it decreases with decreasing $t_{OX}$, $S$ can take ~200 mV/decade for $t_{OX}$ = 2 nm, implying a relatively small off-current. Note that $S$ can be reduced remarkably by increasing $\phi^{SB}$, although there exists a tradeoff between $I_D^P$ and $S$ because $I_D^P$ decreased with increasing $\phi^{SB}$.

Since the spin MOSFET presented here has the excellent performance in sub-100 nm regime and it has a simple structure as shown in Fig. 1(a), one can expect the scaling merits by downsizing the spin MOSFET and high degree of integration. Therefore, the spin MOSFET satisfies all the requirements (i)-(v) for spintronic integrated circuit applications.

In summary, we have proposed and theoretically analyzed the spin MOSFET with HMF source and drain. The spin MOSFET was shown to have magnetization-dependent-output characteristics, high transconductance, amplification capability, low power-delay product, low off-current, and a simple structure compatible with Si-MOS technology, which are all important for integrated circuit applications. The spin MOSFET can be used as a key device for ultrahigh density nonvolatile memory and reconfigurable logic devices based on novel spintronic concepts.

Acknowledgement
This work was supported by the PRESTO program of JST, a Giant-in-Aid for Science Research on the Priority Area "Semiconductor Nanospintronics" (14076207), the IT program of RR2002 from MEXT, and Toray Science Foundation.